# Temperature-Dependent THz Properties and Emission of Organic Crystal BNA

**SAMIRA MANSOURZADEH,**[1, 3, *] **TIM VOGEL,**[1, 3] **MOSTAFA SHALABY,**[2] **AND CLARA J. SARACENO**[1]

[1]*Photonics and Ultrafast Laser Science, Ruhr-University Bochum, Bochum, Germany*
[2]*Swiss Terahertz Research-Zurich, Technopark, Zurich, Switzerland*
[3]*These authors contributed equally to this work.*

*\*mansourzadeh.samira@ruhr-uni-bochum.de*

**Abstract:** As high-average power ultrafast lasers become increasingly available for nonlinear conversion, the temperature dependence of the material properties of nonlinear crystals becomes increasingly relevant. Here, we present temperature-dependent THz complex refractive index measurements of the organic crystal BNA over a wide range of temperatures from 300 K down to 80 K for THz frequencies up to 4 THz for the first time. Our measurements show that whereas the temperature-dependent refractive index has only minor deviation from room temperature values, the temperature-dependent absorption coefficient decreases at low temperature (−24% from 300 K to 80K). We additionally compare these measurements with conversion efficiency and spectra observed during THz generation experiments using the same crystal actively cooled in the same temperature range, using an ultrafast Yb-laser for excitation. Surprisingly, the damage threshold of the material does not improve significantly upon active cooling, pointing to a nonlinear absorption mechanism being responsible for damage. However, we observe a significant increase in THz yield (+23%) at lower temperatures, which is most likely due to the reduced THz absorption. These first findings will be useful for future designs of high-average power pumped organic-crystal based THz-TDS systems.

## 1. Introduction

Terahertz time domain spectroscopy (THz-TDS) systems have become ubiquitous in different fields of fundamental science and technology [1–4]. For many of these applications, ultra-broad THz bandwidths >>3 THz are desired. Such broadband THz emission can be generated using a variety of techniques, including two-color plasma filaments [5], optical rectification (OR) [6,7], and spintronic THz emitters [8,9]. In this respect, OR in organic crystals such as DAST, DSTMS, OH1, HMQ-TMS, and BNA provide a promising platform to generate broadband and efficient THz radiation in a simple collinear geometry. Due to the lower dispersion of the refractive index, organic crystals can be phase-matched in much broader THz bandwidths compared to inorganic crystals. Moreover, they provide high optical-to-THz conversion efficiency in the percent level due to their high nonlinear coefficient [10]. Using the organic crystal OH1, a conversion efficiency of 3.2% was obtained at a low repetition rate of 10 Hz [11]. By pumping a DAST crystal with mid-IR pulses (3.9 µm) Gollner et al. achieved a record value of 6% conversion efficiency at a repetition rate of 20 Hz [7]. Using BNA (N-benzyl-2-methyl-4-nitroaniline) at pump wavelengths between 1150 nm to 1250 nm, results in a conversion efficiency of 0.8% at a repetition rate of 1 kHz [12].

One critical milestone with these materials is to translate this high efficiency to systems with high repetition rates (>>100 kHz), to improve dynamic range and measurement times in spectroscopy experiments. This means at a given input pulse energy, the crystals need to withstand higher average power, without major changes in its properties. Until very recently, this regime of high average power excitation was not explored and mostly low repetition rate pump lasers with a high pulse energy and moderate average power were used, with large excitation spot sizes. However, in the last few years, these crystals have started to be explored for high repetition rate regimes; for example HMQ-TMS [13] and BNA [14] have shown the first promising results with a THz average power of around 1.38 mW and 0.95 mW, respectively, both excited at >10 MHz repetition rate. More recently, we also demonstrated high-power operation of BNA together with high-fields at 540 kHz, resulting in 5.6 mW of THz average power [15]. In [14], we show that the thermal limitations due to high repetition rate can be mitigated by operating in burst mode, thus matching the off-time of the laser with the thermal relaxation time of the crystal. To fully understand the observed limitations, in particular in these new excitation conditions (small spot sizes, high average power) quantitative studies on temperature-dependent THz properties of organic crystals are required. In [16], THz generation in DAST was investigated at different temperatures of 15 K, 54 K and 295 K. They did not observe any enhancement on the THz generation efficiency, however, the position of the absorption peaks in the spectrum shifted. OH1 was also tested at room temperature and 10 K in [17]. They showed, that the efficiency can be increased by 10% through cryogenic cooling. Moreover, a shift in the THz pulse spectrum towards the higher frequency range during crystal cooling was observed.

Concerning the organic crystal BNA, which is most relevant here because of its favorable velocity matching properties at 1030nm, the effect of temperature is expected to be different from other THz organic crystals. In fact, BNA has a melting point of about 125°C while other crystals previously studied are about double this temperature. This is expected to make this material very sensitive to thermal effects. However, to the best of our knowledge, no attempt was made so far to actively control the temperature of BNA as a path to understand its thermal properties. In particular, the investigation of the phase-matching conditions at cryogenic temperatures is critical for understanding efficiency limitations and damage limits. Moreover, linear THz absorption is one of the well-known crucial obstacles to achieve high efficiency and broadband THz radiation in other nonlinear crystals. Therefore, a reduction in THz absorption due to actively cooled crystal can increase the emitted THz strength [18,19].

Here, we explore cryogenic cooling of the organic crystal BNA as a potential path for power upscaling of broadband, high-power THz sources. We show measurements of temperature-dependent refractive index and absorption of BNA at THz frequencies up to 4 THz in a temperature range from 300 K down to 80 K, using a commercial THz-TDS, which to the best of our knowledge have so far not been reported. Based on our measurements, we then discuss the resulting effects on outcoupled THz strength. Finally, we perform first cryogenically cooled OR experiments in BNA based on excitation with our home-built Yb-doped thin-disk laser, operating at ∼13 MHz repetition rate. Our data shows increased THz field strength (+23%), in agreement with the reduced THz absorption at low temperatures. Surprisingly, cryogenically cooling the crystals did not improve the damage threshold of the crystals pointing to nonlinear absorption being the main damage mechanism. Our measurements will be of interest for the THz community to design further high-average power pumped broadband THz-TDS.

## 2. BNA characterization in THz regime

One of the most well-established applications of THz-TDS is their use for the determination of optical properties of materials in the THz frequency range. The goal of all these applications is to measure the frequency-dependent complex refractive index $ñ(ω) = n(ω) − i · k(ω)$ of a sample in THz frequency range. The real term, $n(ω)$, is the refractive index and the imaginary term, $k(ω)$, is the extinction coefficient and ω is angular frequency [20,21].

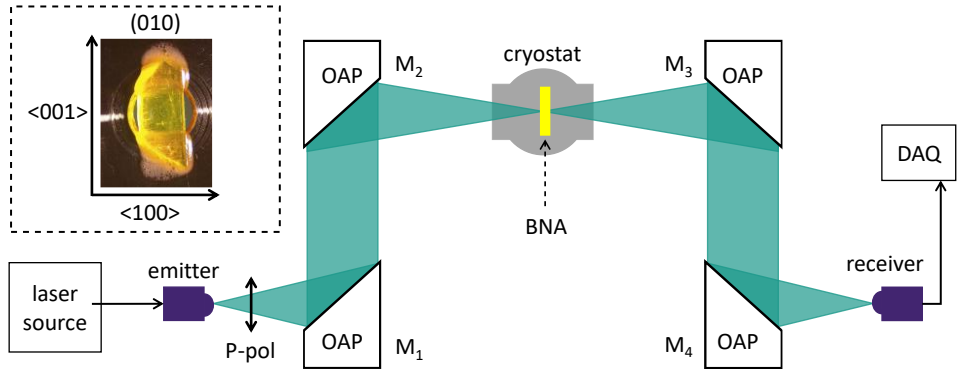

Fig 1. Experimental setup for temperature-dependent BNA characterization in the THz range. Inset shows the crystallographic scheme of BNA together with a picture of the used sample.

The extinction coefficient is related to the absorption coefficient, α, via $k = \alpha c/2\omega$. THz-TDS has the advantage to measure the electric field amplitude and not intensity, which allows to recover the phase of the transmitted signal. The refractive index and absorption coefficient can then be extracted without the Kramer-Kronig approximation [21]. We therefore use THz-TDS in combination with a cryogenically cooled BNA to study its temperature-dependent THz properties.

## 2.1. Experimental setup

In the first experiment, we investigate the refractive index and absorption coefficient of BNA (SwissTHz GmbH) in <001> direction in the THz range using a commercial THz-TDS (Menlo Systems Tera K15). The system provides THz pulses at a repetition rate of 100 MHz, a wide THz spectrum up to 6 THz, and a high peak dynamic range (DR) of >100 dB after averaging over 1000 traces. The full experimental setup is shown in Fig 1. A succession of four off-axis parabolic (OAP) mirrors are employed to collimate and refocus the THz radiation, noted by $M_1$, $M_2$, $M_3$ and $M_4$ in the experimental setup. The first and last OAP ($M_1$ and $M_4$) have a focal length of 50.8 mm. The two middle OAPs ($M_2$ and $M_3$) have focal length of 101.6 mm and provide an in-between focus with full width at half maximum of ~0.5 mm at 1 THz. For the THz transmission measurement, the sample (in this case BNA) is placed inside a cryostat located in the common focal point of $M_2$ and $M_3$.

Besides the cryostat, the other parts of the TDS are in a purged box with a dry nitrogen atmosphere (relative humidity of less than 5%) to prevent absorption of water vapor. To control the crystal temperature, a continuous flow cryostat (Janis ST-100) is used, which can provide temperatures from 300 K down to 80 K using liquid nitrogen cooling and resistive heating. It should be noted that an extended waiting time for each temperature step is necessary, due to the poor thermal conductivity of the BNA crystal. Both entrance and exit cryostat windows are z-cut quartz crystals. All windows are plane-parallel and have a thickness of 3 mm. In order to perform the experiment in a reasonable measurement time and to simplify the processing of the measurement data, the range is set to 60 ps to exclude the reflection of the quartz window of the cryostat. This time window provides a frequency resolution of 16.7 GHz.

BNA is a biaxial crystal and the crystallographic <010> axis is perpendicular to the cleaved facet of the crystal (010). Therefore, the crystallographic <100> and <001> axes lie on the surface of the sample crystal [22]. In this work, the refractive index and absorption coefficients are extracted for <001> axis (see supplementary for <100>). The crystal dimension used in the experiment is 5 mm × 6 mm glued on a copper disk mount with a hole diameter of 4 mm. The glue is an optically transparent, vacuum and cryogenic-temperature compatible product (Masterbond), see inset in Fig 1.

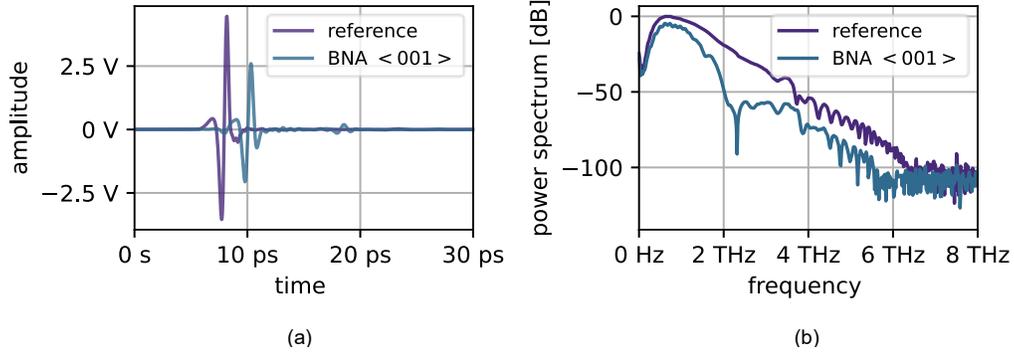

(a)  (b)
Fig 2. Reference and sample THz traces for axis <001> a) in time domain averaged over 10,000 traces. In order to show the structure of the near single cycle THz pulse, the time axis is limited to 30 ps. and b) corresponding frequency domain.

*2.2. Extraction of the refractive index and the absorption coefficient*

As conventional spectroscopic experiments, a complete set of data consists of at least two measurements: first, the THz pulse after propagating through the BNA crystal and second, a measurement in a reference medium under identical conditions but without the sample.

In this experiment, 10,000 single traces are acquired for each temperature for reference and sample. Due to the integrated, fast oscillating delay line of the commercial THz-TDS, one dataset is finished in approximately 450 s. Using single traces gives an advantage over an already averaged THz trace, because statistics and further processing steps can be applied. A pre-processing software, Correct@TDS (developed in the group of Dr. Romain Peretti, Terahertz Photonics Group @ IEMN - CNRS (UMR 8520), publication in preparation), is used to fit specific correction parameters for the delay, dilatation, amplitude noise and periodic sampling. The last error can introduce copies of the original spectra at higher frequencies, when the position signal from the delay line is affected by a periodic error [23]. After this pre-processing, we have two averaged time traces for reference and sample. The respective power spectra show a noise-floor of >100dB. As an example, Fig. 2 shows the reference and sample measurement for the <001> axis at 300 K in time and frequency domain.

To extract the refractive index and absorption coefficient, we developed our own, home-build software called **phoeniks** (**P**ULS **h**ands-**o**n **e**xtraction of $\mathbf{n - i \cdot k's}$) and release it as free and open-source software on Github [24]. It is based on the approach from Pupeza et al. [25], where an artificial transfer-function is modeled with the Fresnel coefficients for reflection and transmission of the THz beam. Additionally, the limited number of echoes from the sample on the THz trace is considered. The echoes are limited due to the finite time window. The numerical transfer function from the experimental data, meaning the division of the sample and reference spectrum in frequency domain, is compared to the artificial transfer function at each frequency point. An optimization algorithm, Nelder-Mead [26], solves numerically for each frequency the inverse problem, finding an optimal $n$ and $k$ which reduce the error function.

Due to the spectrum's complex structure and multiple absorption dips, the phase fluctuates strongly, and a frequency resolution of 1 GHz is necessary, or the algorithm fails to follow the phase. Since the resolution of the native dataset is limited, we apply a window function and zero-padding to the time traces to reach the frequency resolution of 1 GHz.

The thickness of the sample is the only parameter, besides the reference and sample trace, which needs to be supplied to the software and the extraction of the refractive index is highly sensitive to deviations of this parameter. Therefore, the thickness is measured multiple times at different locations of the crystal by a dial indicator (Mitutoyo 543-561D, repeatability of 1 µm) and cross-checked with the total variation method [27]. The dial indicator shows that the crystals is not perfectly flat but has a standard deviation of approximately 32 µm in thickness. The total variation method gives for the dataset not always a parabolic shape with a clear minimum, which makes it challenging to determine the precise thickness.

Due to the temperature sweep, the thickness of the crystal could also be affected. There is no uniaxial thermal expansion coefficient published for BNA, but a recent study about crystalline organic compounds of the Cambridge Structural Database (CSD) suggests, that the mean uniaxial thermal expansion coefficient is $71.4 \times 10^{-6}$ $K^{-1}$[28]. For the tested temperature range, this would correspond to a length change of approximately 1.5%. This study also showed that many crystals contain at least one axis with negative thermal expansion. Due to the limited knowledge of the temperature coefficients of BNA and the likely low influence on the measured data, we neglect the thermal expansion of the crystal in the extraction. For all temperatures, a constant thickness is assumed. We note that in future studies, measuring this coefficient could be highly relevant for a complete picture. To make sure that we get reliable results, we compared it with other software for the extraction of the complex refractive index shown in the supplementary which indicates a very good agreement.

*2.3 Results and discussion*

The extracted refractive index of BNA with a thickness of 487 µm in <001> direction is shown Fig.3 a) for a frequency range up to 4 THz. Extracting data for frequencies higher than 4 THz is challenging: The higher frequency components with a smaller focus scatter by the crystal roughness more than lower frequencies with a larger focus. Moreover, much lower input THz signal strength at higher frequencies compared to the lower frequencies results in low dynamic range which limits the extraction method.

For all frequencies, the value of refractive index at 80 K is higher than the value at 300 K. It shows minor deviations from room temperature values (below 2.1 THz). However, in order to evaluate precisely how the refractive index variation affects the phase-matching condition in OR, we need to calculate the temperature dependent coherence length (effective interaction length in OR). Unfortunately, there are no reported values for the refractive index of BNA in the near infrared range (particularly for 1030 nm) at cryogenic temperatures. Therefore, it is possible to calculate the coherence length of the crystal for different temperatures only by considering the refractive index changes in the THz regime and assuming constant refractive index (room temperature [29]) at the pump wavelength over the varied temperatures. Fig.3 c) illustrates the results. As it can be seen, the coherence length decreases with increasing frequency, and the temperature has only a minimal effect.

The big refractive index jump in vicinity of 2.1 THz for temperatures <120 K (100 K and 80 K) in Fig.3 a) is due to the strong absorption peak occurring at 120 K, see Fig.3 b). The amplitude of this peak is close to the largest absorption coefficient that can be measured reliably by our THz-TDS at this frequency. This calculated value at 2.1 THz based on [30] depends on the dynamic range and is $\alpha_{max} \approx 299$ $cm^{-1}$. The optimization algorithm in the retrieval process minimizes the error between the real and artificial transfer function (see supplementary material). When the absorption coefficient reaches or exceeds $\alpha_{max}$, and therefore the signal drops below the noise floor, the phase information is lost. Consequently, the optimization algorithm will find an arbitrary local minimum which is not necessarily the true refractive index, but just close to the starting position of the algorithm (which is based on the refractive index of the previous frequency point). It affects the refractive index extraction algorithm leading to some uncertainties about refractive index above 2.1 THz at temperatures below 120 K.

Fig.3 b) shows the extracted absorption coefficient. At 300 K, the first strong resonance at 1.57 THz has a linewidth of 95 GHz. By cooling down the crystal to 80 K, the frequency of the resonance has a blue shift to 1.65 THz and the linewidth of the resonance reduces to 45.5 GHz. The second strong resonance appears at 2.14 THz at 300 K, and it shifts to 2.19 THz at 80 K. The reduction of linewidth in this resonance frequency by cooling the crystal is significant: at room temperature, it is 184 GHz and at 80 K it is 58 GHz. The blue shift of the absorption peaks occurs at all resonances. This data shows that particularly at frequencies higher than 1 THz, the reduction of the absorption is very significant. To quantify this reduction, the frequency resolved THz absorption is averaged over the whole frequency range (up to 4 THz) for each temperature. All absorption coefficients are summed up and divided by the number of elements (arithmetic mean). By analyzing this averaged absorption coefficient versus temperature, we observe a reduction by -24% when the crystal is cooled from 300 K down to 80 K.

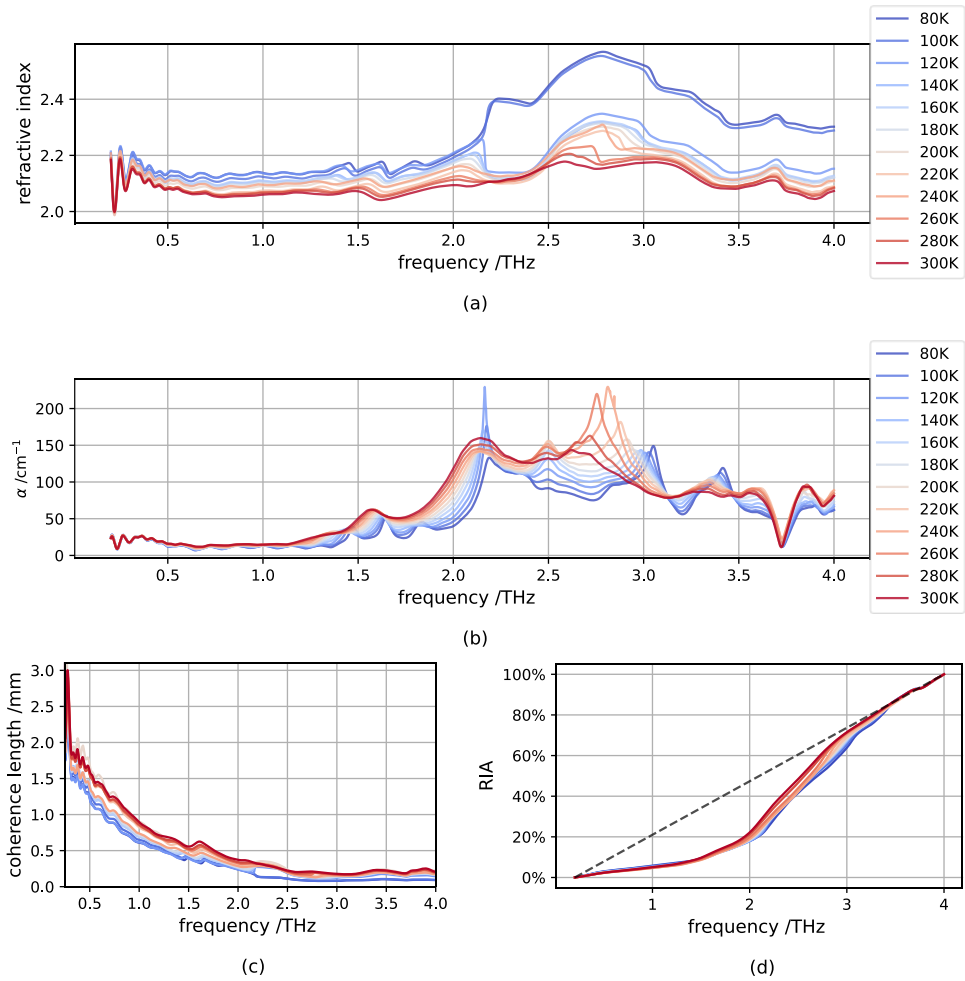

Fig 3. Extracted properties for BNA in <001> direction for different temperatures using phoeniks up to 4 THz. a) refractive index. b) absorption coefficient. c) calculated coherence length with constant refractive index for the pump wavelength (1030 nm). d) relative integrated absorption showing a slower increase of RIA for lower temperatures.

Fig.3 d) shows the relative integrated absorption (RIA) in the investigated THz region. The integrated absorption value highlights the relative change of absorption at each temperature value independent of the absolute absorption. The black, dashed line stands for a constant absorption and acts as a guide for the eye. In the case of no absorption, the curve should be horizontal from this specific frequency on. At all temperatures, the absorption rises very slowly up to 2 THz, where a stronger slope starts to set in. Here, the relative difference between the temperatures is the largest, and higher temperatures leading to a faster rise in absorption. At 3.5 THz, the relative, integrated absorption rises with near maximum rate.

## 3 THz generation in temperature-controlled BNA crystal

### 3.1 Experimental setup

In this section, we present THz generation experiments based on OR using an actively cooled BNA crystal pumped with a high-average power Yb-based thin-disk oscillator. The main aim of these experiments is to correlate the observed temperature-dependent trends with the measured THz properties of the crystals in the previous section.

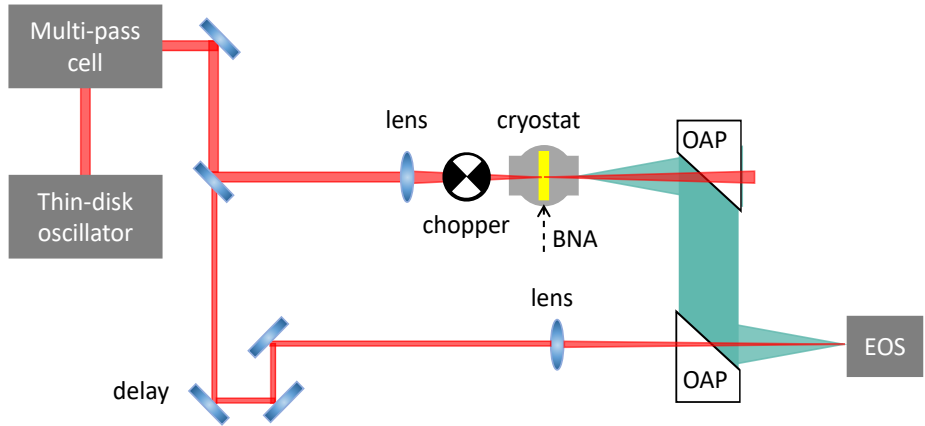

Fig 4. Experimental setup for THz generation based on OR in temperature-controlled BNA.

The full experimental setup is shown in Fig 4. The laser pulse has a temporal duration of 550 fs and a central wavelength of 1030 nm. The pulse is externally compressed down to 80 fs using a Herriot-type multi-pass cell. More details about the laser and compression setup can be found in [31]. After the compression stage, the laser beam is guided towards the THz-TDS and splits in two parts: more than 98% goes to the pump arm to pump the same BNA crystal as used in the first experiment. The remaining 2% go to the probe arm of the THz-TDS and sample the THz trace in a standard electro-optic sampling (EOS) setup. The pump beam diameter at the position of the BNA crystal is ~0.36 mm ($1/e^2$) provided by a focusing lens with a focal length of 200 mm. The generated THz beam is collimated and refocused on the detection crystal using two identical OAPs with a diameter of 76.2 mm and a focal length of 152.4 mm. The detection crystal is a 0.5 mm GaP placed in the focus of the second OAP. Further reduction of the thermal load on the BNA crystal is obtained by operating in burst mode by using an optical chopper with a duty cycle of 50% and a chopping frequency of 2.7 kHz is placed before the crystal. The EOS data is acquired using a lock-in amplifier, which records the signal from the balanced photodetector and the digitized position of the delay line in the probe arm. The modulation frequency of the pump beam (from the previously mentioned optical chopper) is used as a reference for the lock-in amplifier. The delay line is a shaking retroreflector with a sinusoidal movement with a traveling range of 13 ps and a shaking frequency of 5 Hz.

### 2.3 Results and discussion

Fig.5 a) shows the detected THz trace by EOS in the time domain for different temperatures. In all cases, the crystal is pumped with a fixed laser power of 1.3 W. All EOS traces are recorded in 78 s and averaged over 390 traces. Fig.5 b) combines the findings from both experiments including characterization and generation time domain results. The turquoise curve shows the calculated average absorption up to 4 THz for each temperature using Fig.3 b). The purple curve shows the peak-peak electric field from the generated THz radiation in time domain taken from Fig.5 a). This curve verifies the enhancement of THz conversion efficiency by cooling down BNA. The influence of the THz absorption reduction results in a peak-peak electric field rise with +23% (from 280 K to 80 K) at same excitation conditions which is in a good agreement with the reduction of THz absorption by −24% obtained from purple curve. We note here that BNA has a poor thermal conductivity, therefore the cooling finger temperature may not be representative of the actual temperature of the crystal when applying the laser power on it. The real temperature will certainly be higher at the point of highest intensity where the laser is applied, thus the improvements observed are moderate. However, the observed trend is clear and points us to future improvements for example using actively controlled heatsinked BNA crystals.

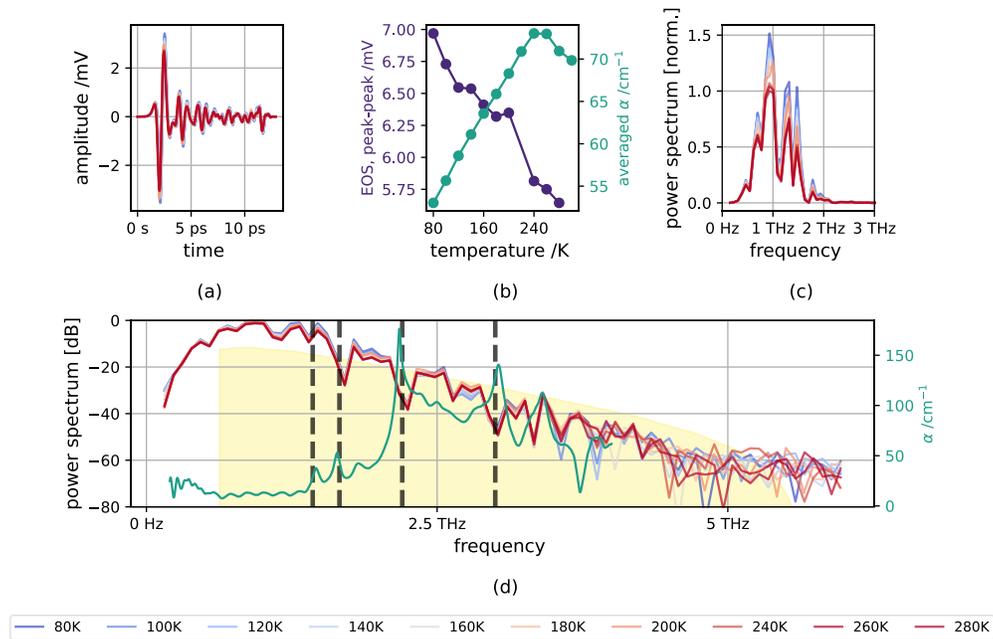

Fig 5. THz generation via OR in BNA pumped with a 1030-nm ultrafast laser. a) THz trace in time domain averaged over 390 traces in 78 s for all temperatures. b) turquoise curve: reduction of the average absorption when the temperature is reduced. Purple curve: Consequently, the peak-peak electric field from the generated THz radiation increases when the crystal is cooled. c) Power spectrum on the linear scale, all curves are normalized to the peak value at 300 K. d) Corresponding spectrum in frequency domain. Turquoise curve shows the absorption curve at 80 K obtained from the previous section. Shaded yellow area shows response function of 0.5 mm GaP.

Corresponding spectra are shown in Fig.5 c) using Fourier transformation for all temperatures. In order to show the change of the strength at each frequency components by varying the temperature, the power spectra are shown in a linear scale. The power enhancement is significant for the frequencies up to 2 THz. Compared to room temperature, the power at approximately 0.9 THz at 80 K increases even more than 50%.

To give a more detailed view, the same spectra are shown in a logarithmic scale in Fig.5 d). Since the effect of cooling is stronger at higher frequencies, it is important to evaluate the factors which limits the detected spectrum shape. Therefore, we introduce the response function of 0.5 mm GaP calculated according to the reference [32], the low-pass filtering of the lock-in amplifier, low-/high-pass filtering of OAPs, simulated on the model from reference [33] and the transmission function of the PTFE filter [34] used to suppress the residual pump beam. All these effects are combined and shown as a yellow shaded area. As it can be seen, even though the low-pass filtering from the 3 mm hole in the OAP is negligible, the combined low-pass filtering effects slightly limit us to observe higher THz frequencies than 5 THz. In order to show if the dips in the detected spectrum fits with the peaks in the absorption curve, the absorption curve for 80 K is plotted as turquoise solid line. The positions of the dips are in very good agreement with the position of the absorption peaks. It should be noted that the frequency resolution of the generated spectrum is 77 GHz which is 5 times worse than the resolution in absorption peak curve. This might result to slight mismatch between two curves. Moreover, the detected spectra are recorded in lab environment (unpurged) where water vapor modulates the THz spectrum with additional dips. Nevertheless, the influence of the measured absorption and their change when changing the crystal temperature are clearly visible.

Surprisingly, our observation is that the maximum pump power applied to the crystal before irreversible damage changes only very slightly by reducing the temperature. The maximum applicable intensity with the chopper with duty cycle of 50% at 280 K is 2.1 $GW/cm^2$ and at 80 K is 2.6 $GW/cm^2$. At first glance, these results indicate that a nonlinear absorption mechanism is most likely mainly responsible for damage in the BNA crystal, and linear absorption seems to have only a minor influence. However, a more detailed exploration is required for this. In order to disentangle linear and nonlinear absorption effects, additional experiments to precisely identify the physical mechanism of the generated heat are required. In fact, in the case of high average powers and high repetition rates, a significant amount of heat can be generated due to pump laser absorption. Therefore, it is required to quantify the generated heat due to these absorptions in organic crystals, for example in the same way as it was done in [19], using a cryostat and a temperature controller. In order to distinguish between linear and nonlinear absorption, the experiment needs to be conducted with a continuous wave and an ultrafast pulsed laser. Furthermore, we mentioned above that the actual temperature rise of the crystal due to the laser when actively cooled will be inevitably higher than that of the cooling finger in the cryostat, this needs to be additionally quantified to test the expected change in damage threshold. Additional temperature resolved experiments, such as a study on the influence on pulse picking, burst length and single pulse energy change, as well as possibly pump-probe experiments on the crystal might be helpful too to evaluate the different physical effects affecting OR.

## 5   Conclusion and outlook

We present an investigation of cryogenic cooling of the organic crystal BNA as a potential path for power upscaling of broadband high-power THz sources. We measure the temperature-dependent refractive index and absorption of BNA at THz frequencies up to 4 THz in a temperature range from 80 K up to 300 K using a commercial THz-TDS. Our data shows negligible change in refractive index (with respect to optical rectification), but a significant reduction of THz absorption by −24% by cooling the crystal from room temperature down to 80 K. The larger temperature dependence is observed at frequencies >2 THz. Moreover, we perform first cryogenically cooled OR experiments in BNA pumped with our home-built thin-disk laser, operating at ∼13 MHz repetition rate to correlate the observed trends with the measured crystal properties. We observe an increased THz field strength (+23%), clearly corresponding to a reduced THz absorption at low temperatures, providing one important ingredient for future optimization of high-power broadband THz sources.

Our findings show that cryogenic cooling is a straightforward route to enhance the THz efficiency and power spectral density at given high frequencies by reducing the absorption in organic crystals. To the best of our knowledge, these are the first reported temperature-dependent THz refractive index and absorption measurements for any organic crystal; and further the first time these are correlated with THz emission data. We believe these measurements will be highly relevant for the THz community, as high-power pumping becomes increasingly popular. In future studies, we plan to investigate the temperature dependent refractive index of BNA at NIR regime to be able to estimate precisely the effect of the cooling on the coherence length. Moreover, changing the detection parameters and crystal to be able to detect the higher bandwidth where the effect of the cooling should be stronger. Investigating of other organic crystals such as MNA and NMBA in THz and NIR regime will be the next step to study the temperature-dependent material properties at THz regime.

**Acknowledgements.** This project was funded by the Deutsche Forschungsgemeinschaft (DFG) under Germany's Excellence Strategy - EXC 2033 - 390677874 – RESOLV and also under Project-ID 287022738 TRR 196 (SFB/TRR MARIE). These results are part of a project that has received funding from the European Research Council (ERC) under the European Union's Horizon 2020 research and innovation programme (grant agreement No. 805202 - Project Teraqua). The project "terahertz.NRW" is receiving funding from the programme "Netzwerke 2021", an initiative of the Ministry of Culture and Science of the State of Northrhine Westphalia. The sole responsibility for the content of this publication lies with the authors.
We thank Romain Peretti (University of Lille), Ioachim Pupeza (Max-Planck Institute for Quantum Optics), Andrew Burnett (University of Leeds), and Milan Öri (Menlo Systems GmbH) for their fruitful discussions.

**Data availability.** The data of this paper can be found under https://doi.org/10.5281/zenodo.7857371

**Disclosure.** M.S. is an employee of Swiss Terahertz GmbH.

**Supplemental document.** See Supplement 1 for supporting content.